\documentclass[prc,aps,showpacs,twocolumn,nofootinbib,superscriptaddress,floatfix]{revtex4-1}

\usepackage{graphicx}
\usepackage{epsfig}
\usepackage{amsmath,amssymb}
\usepackage{amsmath,amssymb}
\usepackage{bm}
\usepackage{color}

\begin{document}

\title{Information content of the low-energy electric dipole strength: correlation analysis}

\author{P.--G. Reinhard}
\affiliation{Institut f\"ur Theoretische Physik II, Universit\"at
Erlangen-N\"urnberg, Staudtstrasse 7, D-91058 Erlangen, Germany}

\author{W. Nazarewicz}
\affiliation{Department of Physics \&
Astronomy, University of Tennessee, Knoxville, Tennessee 37996, USA}
\affiliation{Physics Division, Oak Ridge National Laboratory, Oak Ridge, Tennessee 37831, USA}
\affiliation{Institute of Theoretical Physics, University of Warsaw, ul. Ho\.za
69, 00-681 Warsaw, Poland}

\date{\today}

\begin{abstract}
\begin{description}
\item[Background]
Recent experiments on the electric dipole (E1) polarizability in heavy nuclei have stimulated  theoretical interest in the low-energy electric dipole strength, both isovector and isoscalar.
\item[Purpose]
We study the information content carried by the electric dipole   strength 
with respect to isovector and isoscalar indicators characterizing bulk nuclear matter and finite nuclei. To separate isoscalar and  isovector modes, and low-energy strength and giant resonances, we analyze the E1 strength
as a function of 
excitation energy $E$  and  momentum transfer $q$. 
\item[Methods]
We  use the self-consistent nuclear density functional theory with Skyrme energy density functionals, augmented by the random phase approximation, to compute the E1 strength, and covariance analysis to assess correlations between observables.
Calculations are performed for spherical,  doubly-magic nuclei $^{208}$Pb and $^{132}$Sn.
\item[Results] 
We demonstrate that E1 transition densities in the low-energy region
below the giant dipole resonance  exhibit  appreciable  state dependence and multi-nodal
structures, which are fingerprints of weak
collectivity. The correlation between the accumulated  low-energy strength and symmetry energy is weak, and dramatically depends on the energy cutoff assumed.
On the other hand, a strong correlation is predicted between isovector indicators and the accumulated  isovector strength at $E$ around 20\,MeV and  momentum transfer $q\sim 0.65$\,fm$^{-1}$.
\item[Conclusions]
Momentum- and coordinate-space pattern of the low-energy dipole
  modes indicate a strong fragmentation into  individual particle-hole
  excitations. The global measure of
 low-energy dipole strength poorly correlates with the nuclear symmetry energy and other isovector characteristics. Consequently, our results do not support the suggestion that there exists  a collective  ``pygmy dipole resonance," which is a strong  indicator of nuclear isovector properties. By considering nonzero values of momentum transfer, one can isolate individual excitations and nicely separate low-energy excitations  from the $T=1$ and $T=0$ giant collective modes. That is,  measurements at $q >0$ may serve as a tool to correlate the E1 strength with certain bulk observables, such as incompressibility or symmetry energy.
\end{description}
\end{abstract}

\maketitle

\section{Introduction}

The   electric dipole response  of the atomic nucleus carries fundamental information on bulk nuclear properties and shell structure 
\cite{(Boh75),(Rin00),(Har01),(Lip82),(Lip89),(Rei99),(Sat06),(Paa07),(Tri08)} and it also plays an important  role in 
nuclear reactions involving photo-nuclear processes. In particular, E1 strength
impacts  photo-absorption and radiative particle-capture
processes occurring in stars during the cosmic nucleosynthesis \cite{(Gor98),*(Gor04),*(Dao12)}
and the transmutation of radioactive nuclear waste \cite{(Erh10), *(Bea12)}.

The significance of the E1 strength, especially in the context of neutron-rich nuclei, has led to  appreciable experimental progress in this area
\cite{(Rye02),(Adr05),(Zil02),*(Har04),*(Sav06),*(Sav08),*(End10),*(Sav11),*(Isa11),*(End12),(Sch08),*(Mas12),(Kli07),(Ton10),(Car10),*(Wie09),(Tam11)}.
Much excitement has been brought by a suggestion that the low-energy E1 strength,
dubbed ``Pygmy Dipole Resonance" (PDR)  \cite{(Brz68),(Lan71)}, is collective in nature, and can be understood as a motion of skin neutrons against the proton-neutron core. 

The current theoretical situation regarding the collectivity of the low-energy E1 strength is fairly confusing. While some papers advocate the existence  of a collective PDR mode \cite{(Pie06),(Kli07),(Tso08),*(Tso04),(Car10),(Pie11),(Bar12)}, having strength that is correlated  with the nuclear symmetry energy \cite{(Pie06),(Kli07),(Tso08),*(Tso04)}, the collectivity of the low-energy E1 transitions, and their relevance to isovector nuclear matter properties (NMP),
have been questioned by several other studies
\cite{(Rei10),(Gam11),(Dao11),(Ina11),(Roc12),(Yuk12)}.

In particular, the  covariance analysis of Ref.~\cite{(Rei10)} has suggested the lack of correlation between PDR strength and nuclear isovector properties, such as neutron-skin 
thickness $r_{\rm skin}$ 
and  symmetry energy. One reason for this was attributed to the local single-particle structure around the Fermi level 
(i.e., shell effects),  which vary 
rapidly with global nuclear matter characteristics. 
On the other hand, the dipole polarizability $\alpha_\mathrm{D}$
-- a global measure of the E1 strength -- seems to be an excellent isovector indicator  
\cite{(Rei10),(Tam11),(Pie12),(Vre12)}.
Similar conclusions have been reached in Ref.~\cite{(Dao11)}, which concluded that the current data on the low-energy strength, and theoretical predictions,  are both too uncertain for a quantitative analysis.

The aim of this work is to clarify the situation by means of the
stringent correlation analysis based on methods of data analysis
  using least-squares optimization. We will consider, in particular,
  the correlations between various nuclear observables (including NMP and properties of finite nuclei)  and E1 strength.
Our paper is organized as follows. Section~\ref{Themodel} describes our nuclear EDF+RPA approach and motivates the choice of EDFs used. The definitions pertaining to E1 form factors, strength functions,  and resulting E1 sum rules are discussed in Sec.~\ref{E1strength}. The dependence of the low-energy E1 strength on selected nuclear matter properties is studied in Sec.~\ref{LEE1strength}.
The results for transition densities are presented in
Sec.~\ref{Trdensities}, and Sec.~\ref{E1correlations} contains the
correlation analysis. 
While most calculations in this study were done for $^{208}$Pb, 
Sec.~\ref{Sncase} contains illustrative examples for $^{132}$Sn that demonstrate that our findings are general.
Finally, Sec.~\ref{Conclusions} contains
conclusions of the work.

\section{Model}\label{Themodel}

The present investigation is based on the self-consistent nuclear
density functional theory in the
Skyrme-Hartree-Fock (SHF) variant \cite{(Ben03),(Erl11)}. The form of the SHF energy density functional (EDF) is derived from
rather general arguments of a low-momentum expansion \cite{(Neg72),(Car10a)}.
However, since the coupling constants of EDF cannot be determined precisely
from underlying nuclear forces, they are usually adjusted to
selected nuclear data \cite{(Ben03),(Klu09),(Kor12)}.  We use here
two recent EDF
parameterizations  which try to embrace
a large collection of various nuclear data \cite{(Klu09),(Kor10)}.
The survey \cite{(Klu09)} selected a  pool of spherical nuclei
which have been checked to be well described by a mean-field model
\cite{(Klu08)}. The resulting parameterization  SV-min was optimized by means of a least-squares procedure. Its (parabolic) least-squares landscape
$\chi^2=\chi^2(\mathbf{p})$, where $\mathbf{p}$ stands for the
multitude of SHF parameters, carries important information on the 
uncertainties, in particular on model features  that are weakly determined by the optimization process. This can be exploited by applying the standard covariance analysis as was done, e.g., in Refs.~\cite{(Rei10),(Fat11)}. To estimate the uncertainty related to  the choice of fit observables, we alternatively look at the parameterization UNEDF0 
\cite{(Kor10)}, which covers a much larger set of ground state data, also
including  many deformed nuclei.

It is found that many aspects of EDF  can be characterized by relatively few
NMP, such as incompressibility $K$, isoscalar effective mass $m^*/m$,
symmetry energy $a_\mathrm{sym}$, and Thomas-Reiche-Kuhn (TRK) sum rule
enhancement $\kappa$ (related to the isovector effective mass).  By providing a set of parameterizations for which these NMP are
 systematically varied \cite{(Klu09)}, one can track explicitly the influence of the NMP on various nuclear properties.  Base point of this set is the EDF called SV-bas,  for which four
NMP had been fixed: $K=234$
MeV, $m^*/m=0.9$, $a_\mathrm{sym}=30$ MeV, and $\kappa=0.4$ (chosen
such that the giant dipole resonance (GDR) in $^{208}$Pb is described
correctly). Starting from SV-bas, four sets of parameterizations were
produced \cite{(Klu09)} by systematic variation of each one of the four
NMP. It is to be noted that SV-bas and its systematically
  varied sister parameterizations are adjusted with additional
  constraints on NMP. SV-min, on the other hand, is fitted only to
  data from finite nuclei and thus covers all uncertainty on NMP as
  left open by nuclear ground state data, a feature which make it
  ideally suited for our covariance analysis.

The focus of our survey is on  properties of nuclear excitations  in
  the electric dipole channel.  To compute the E1 strength function, we
employ the random phase approximation (RPA), using  the
same EDF as for the ground state to guarantee the full self-consistency. The details of RPA calculations
follow Ref.~\cite{(Rei92a),(Rei92b)}.

\section{Strength functions}\label{E1strength}

In the following, we discuss various aspects of the electric dipole strength, both isoscalar ($T=0$) and isovector ($T=1$).  Detailed information on the strength is  contained in  the dipole transition form factor:
\begin{equation}
  {F}^{(T)}_\nu(q)
  =
  \langle\Psi_\nu|
   \sum_{\alpha=1}^Aj_1(qr_\alpha)Y_{10}(\Omega_\alpha)\hat{\Pi}_T
  |\Psi_0\rangle
\label{eq:Ftra}
\end{equation}
where $\alpha$ runs over all nucleons,
$\Psi_\nu$ is the wave function of the  $\nu$-th RPA excitation,  $\Psi_0$ the
ground state wave function, $q$ is  momentum transfer,  $T$ is the isospin quantum number, and
$\hat{\Pi}_T$ stands for the isospin projector:
\begin{eqnarray}
  \hat{\Pi}_{0}
  &=&
  \hat{\Pi}_\mathrm{prot}
  \!+\!
  \hat{\Pi}_\mathrm{neut}
  \;,\;
  \hat{\Pi}_{1}
  =
  \frac{N}{A}\hat{\Pi}_\mathrm{prot}
  \!-\!
  \frac{Z}{A}\hat{\Pi}_\mathrm{neut}.
\label{eq:isoproj}
\end{eqnarray}
(For spherical nuclei, such as  $^{208}$Pb and $^{132}$Sn considered in this work, it
is sufficient to consider the single mode with the magnetic quantum
number  $\mu=0$.)
The $q$-dependence carries the pattern of an excitation
  mode. A complementing coordinate-space view is provided by
radial transition density, which is the Fourier transform of ${F}^{(T)}_\nu(q)$: 
\begin{equation}
  \rho_\nu^{(T)}(r)
  =
  4\pi\int_0^\infty dr\,r^2\,j_1(qr)\,F_\nu^{(T)}(q).
\label{eq:rhotra}
\end{equation}

Since the RPA excitation spectrum of a heavy nucleus such as  $^{208}$Pb is fairly dense, it is more convenient 
to consider the energy distribution of form factor strength:
\begin{equation}
  S_F^{(T)}(q,E)
  =
  \frac{1}{q^2}
  \sum_\nu|{F}^{(T)}_\nu(q)|^2
  \mathcal{G}_\Gamma(E-E_\nu),
\label{eq:SFtra}
\end{equation}
where $E_\nu$ is the RPA excitation energy of state $\nu$ and  $\mathcal{G}_\Gamma(E-E_N)$ is
a Gaussian folding function having energy-dependent width 
$
  \Gamma(E_\nu)
  =
  \mbox{max}\left[0.2\,
    \mathrm{MeV},{(E_\nu-8\,\mathrm{MeV})}/{6\,\mathrm{MeV}} \right]
$ --
to simulate nuclear damping effects, which increase with energy.
We note small $\Gamma$ (and improved resolution)   in the low energy region, which is the very focus of this work.

The factor $q^{-2}$ in Eq.~(\ref{eq:SFtra}) is introduced to  cancel  the leading $q^2$-dependence 
of ${F}^{(T)}_\nu(q)$ at
small momentum transfer. Indeed, at 
$q\rightarrow 0$,
the isovector strength $S_F^{(1)}(q,E)$ is proportional to that of
the familiar  dipole operator $\hat{D}=rY_{10}\hat{\Pi}_1$, which we denote by $S_D(E)$. The
next-order term in $q$  leads to the ``compressional dipole'' operator
$\hat{D}_c^{(T)}=(r^3- {5\over 3}\langle r^2\rangle r)Y_{10} \hat{\Pi}_{T}$, the leading observable in the isoscalar channel \cite{(Gia81),(Har81),(Ham98),(Kol00),(Kva11)}, and the 
corresponding strength is denoted by $S_{D_c}^{(T)}(E)$.
For energy cuts at  finite $q$, we shall consider the general dipole operator  $\hat{F}=j_1(qr)Y_{10}\hat{\Pi}_{T}$ as in Eq.~(\ref{eq:Ftra}). 

\begin{figure}[htb]
  \includegraphics[width=0.6\columnwidth]{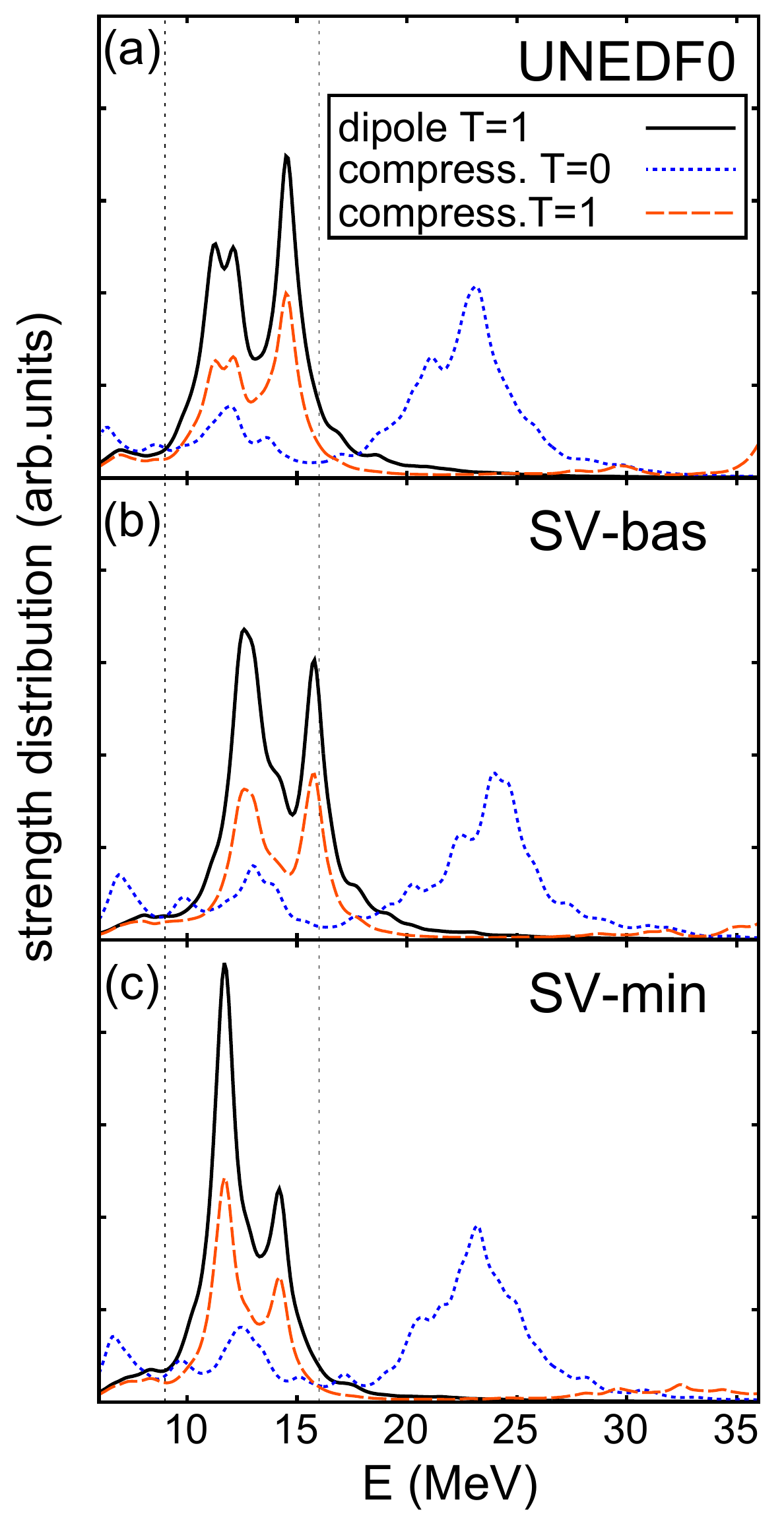}
  \caption[T]{\label{strengthsexamples}
  (Color online) Isoscalar and isovector E1 strength 
   $S_D$ (solid line),  $S_{D_c}^{(0)}$ (dotted line), and  $S_{D_c}^{(1)}$ (dashed line)
  in $^{208}$Pb predicted with (a) UNEDF0, (b) SV-bas, and (c) SV-min EDFs. The region of GDR is marked by vertical dotted lines; these limits are somewhat arbitrary as the GDR strength does not show clear energy bounds.
 }
\end{figure}
To assess the importance of the low-energy E1 strength, we introduce the
accumulated strength:
\begin{equation}
A_n^\mathcal{O}(E)
  =
  \int_0^E dE'\,E'^n\, S_\mathcal{O}(E'), \label{summedS}
\end{equation}
where $\mathcal{O}$ can mean either the mere dipole $D$, compressional dipole $D_c$, or general dipole  $F$  at a given value of $q$. 
In Eq.~(\ref{summedS}), we use a constant folding width $\Gamma=0.5$
MeV in $S_\mathcal{O}$.
For the dipole operator, the inverse-energy-weighted sum rule $A_{-1}^D(\infty)$
is related to the electric dipole polarizability \cite{(Lip89)}
\begin{equation}
\alpha_{\rm D}=
2\sum_{\nu\in\mathrm{RPA}}
(|\langle\Psi_\nu|\hat{D}|\Psi_0\rangle|^2/E_\nu),
\end{equation}
while the energy-weighted sum rule $A_{1}^{D}(\infty)$ is the
TRK sum rule. 

Figure \ref{strengthsexamples} shows 
E1 strength $S_D(E)$ and $S_{D_c}^{(T)}(E)$ in $^{208}$Pb predicted
with UNEDF0, SV-bas, and SV-min EDFs. The isovector dipole  strength $S_D$ is concentrated 
in the GDR region of 12-16 MeV. The systematic shift of the GDR
strength between EDFs reflects the change in TRK sum rule enhancement
($\kappa$=0.4 for SV-bas, 0.25 for UNEDF0, and 0.08 for SV-min) \cite{(Klu09)}. The isovector compressional strength 
$S_{D_c}^{(1)}$ roughly follows $S_D$. It does also contain a
  high energy branch above 30 MeV which, however, is not of interest in the context of this study. The isoscalar  compressional strength is fairly broad, with a low-energy component
at 5-18 MeV and a high-energy concentration at 20-30
MeV \cite{(Paa07)}.

\begin{figure}[htb]
  \includegraphics[width=\columnwidth]{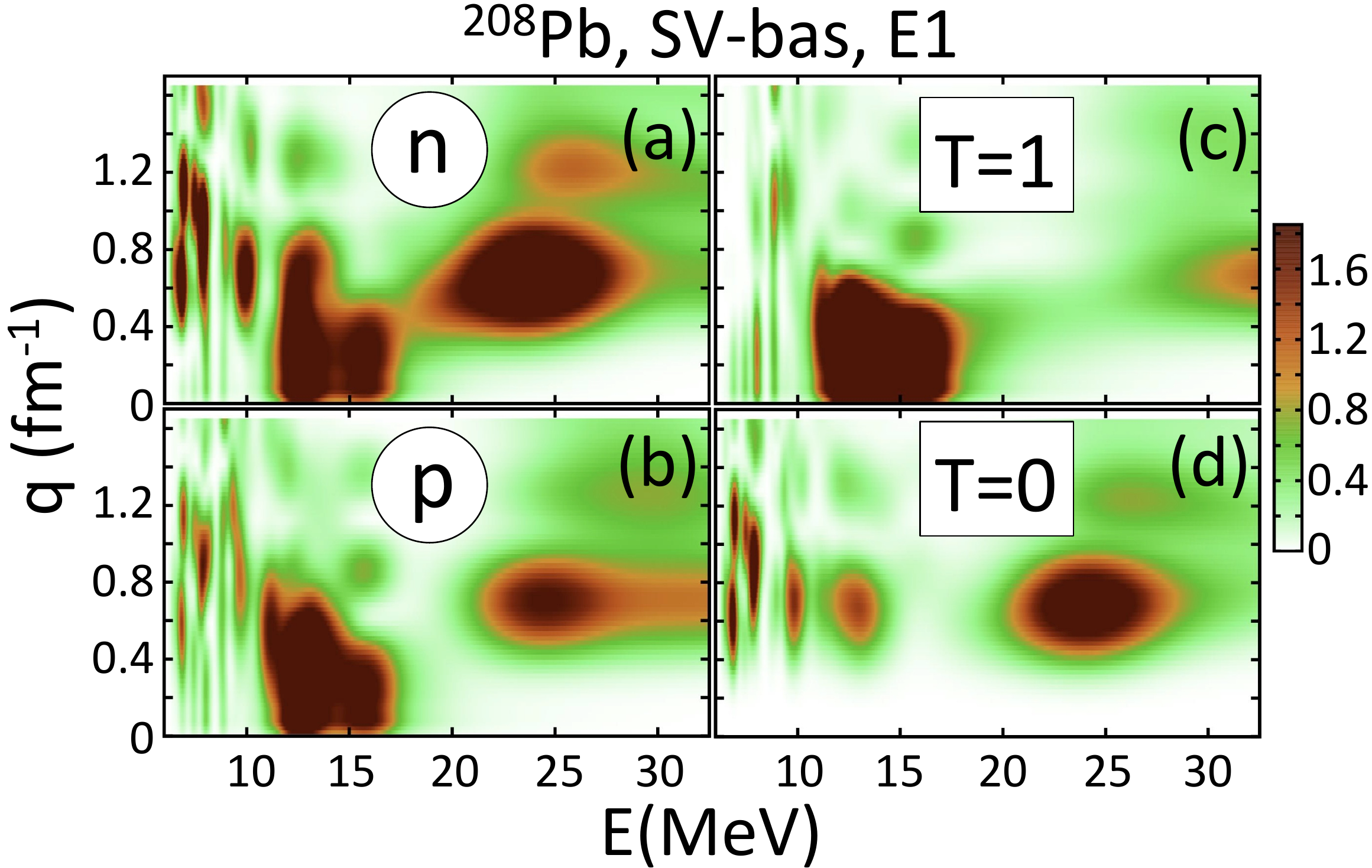}
  \caption[T]{\label{formfactors}
  (Color online) E1 strength (\ref{eq:SFtra})
  for $^{208}$Pb computed with SV-bas as a function of $E$ and $q$. (a) Neutron strength; (b) proton strength; (c)   $T=1$ strength (scaled by 2); and (d) $T=0$ strength (scaled by 0.25).
  Since  $S_F$ rapidly decreases  with  $q$, rather than the actual strength (\ref{eq:SFtra}) we show the $q$-scaled strength
  $  \tilde{S}_F^{(T)}(q,E)
  =
  S_F^{(T)}(q,E)(1+{a_q}q^2)^2$ with $a_q=9$\,fm$^2$. 
 }
\end{figure}

Using the standard representation of E1 strength as shown in Fig.~\ref{strengthsexamples}, it is difficult to separate the low-energy
strength from the tails of giant resonance modes.  A better separation can be obtained by considering nonzero values of $q$, i.e., by analyzing the full distribution  $S_F^{(T)}(q,E)$. Figure~\ref{formfactors} shows the structure of the E1 strength (\ref{eq:SFtra}) -- together with individual proton and neutron
strength distributions --  for SV-bas in the $(E,q)$ plane.
As expected, the isoscalar channel shown in Fig.~\ref{formfactors}(d) has no strength at
$q\rightarrow 0$, as the spurious isoscalar collective center-of-mass E1 mode is removed in our calculations.  The large isoscalar strength appears at somewhat larger momentum transfer around $q=0.65$fm$^{−1}$. Therein,
one can clearly see the two branches of the isoscalar dipole compressional
mode at 6-13 MeV  and 24 MeV.   The  GDR centered around 13 MeV dominates the $T=1$ strength at low values of $q$. 
It seems that the group of excitations between 6 and 11 MeV 
has predominantly isoscalar compressional character. They seem to be higher-$q$ extensions of some low-energy isovector dipole strength (often referred to as PDR strength). These low-energy dipole transitions can thus be considered as shadows of the low-energy dipole compressional modes.
The neutron
and proton E1 strengths  displayed in Figs.~\ref{formfactors}(a) and (b), respectively, show that
the states having large transition form factors are {\it all}  mixed proton and neutron excitations. While the  neutrons
carry more strength generally,  pure neutron modes do not appear.

\begin{figure}[htb]
  \includegraphics[width=\columnwidth]{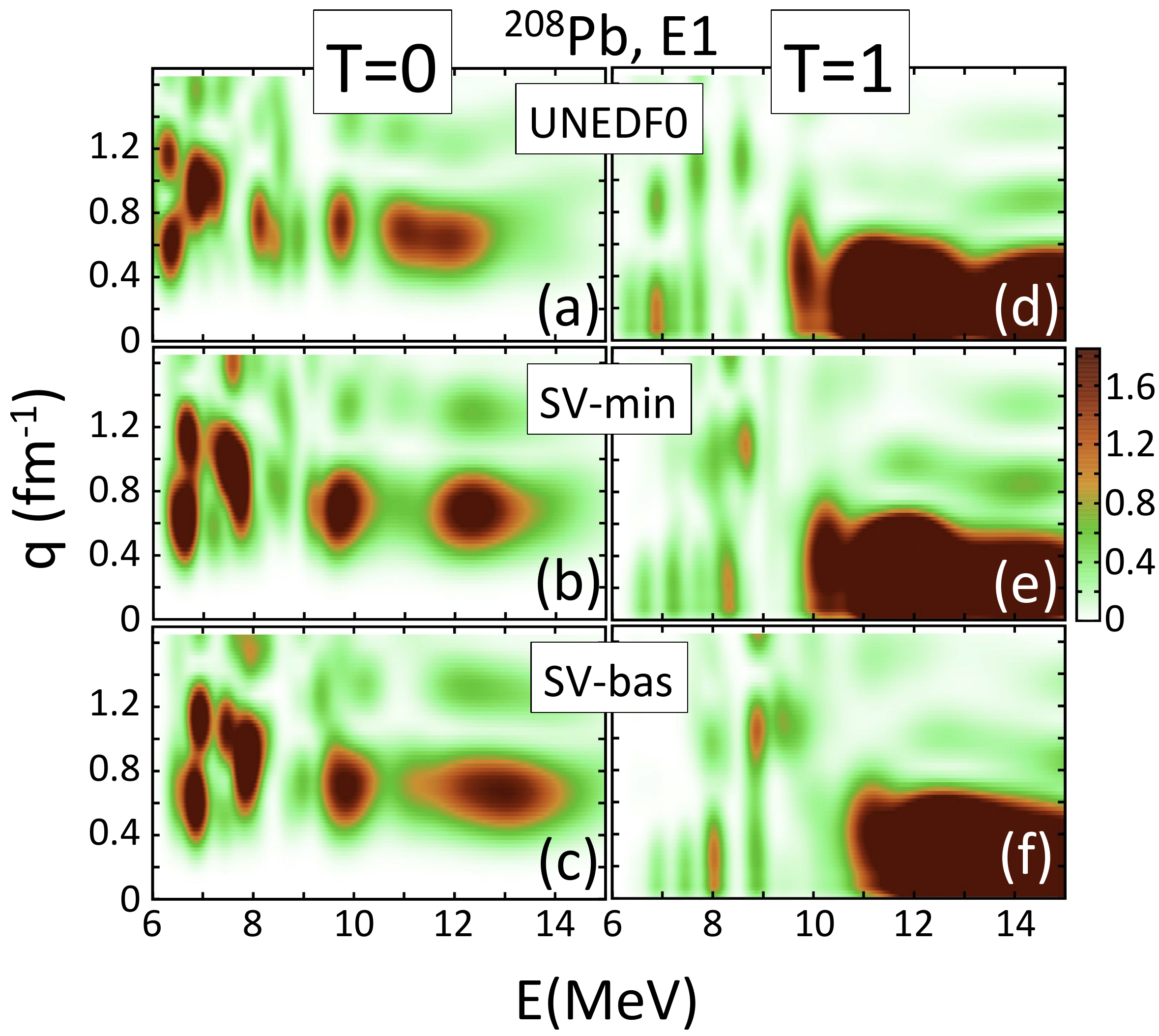}
  \caption[T]{\label{strengthforces}
  (Color online) Similar as in Fig.~\ref{formfactors}   for  $T=0$ (left) and $T=1$ (right) E1 strength computed with (a) UNEDF0, (b) SV-min, and (c) SV-bas EDFs.
 }
\end{figure}

The upper branch of the compressional mode around 22--26 MeV shown in
Fig.~\ref{formfactors} is  not so interesting for
our  study.  Therefore, in the following, we will restrict the energy range.
Figure~\ref{strengthforces} illustrates  the low-energy pattern of the E1 strength
for three EDFs of Fig.~\ref{strengthsexamples}. It is encouraging to see that the general behavior of E1 strength is   EDF-independent, so the detailed discussion of RPA modes will be based on SV-bas calculations.
In the energy range below 10\,MeV, only 
a handful of modes can be seen. Actually, we find more RPA states  
(e.g., 20 modes between 6 and 10 MeV) in that range, but only
a third of them are  visible in E1 excitation.
These strong-E1,  low-$E$ modes have a fairly diverse  character. The lowest
states below 7.5\,MeV are  predominantly isoscalar in character, having some contribution to  the $T=1$ channel.
The mode at 8\,MeV has a mixed character (isovector at low $q$-values and isoscalar at higher $q$), while
the  mode at 9 MeV is predominantly an  isovector excitation.
The states around 10 MeV are again dominated by $T=0$. Their $T=1$ component is weak, and it is mainly concentrated at larger  $q$, i.e., these states do not carry significant dipole ($q=0$) strength.

\section{Transition densities}\label{Trdensities}

\begin{figure}[htb]
  \includegraphics[width=0.8\columnwidth]{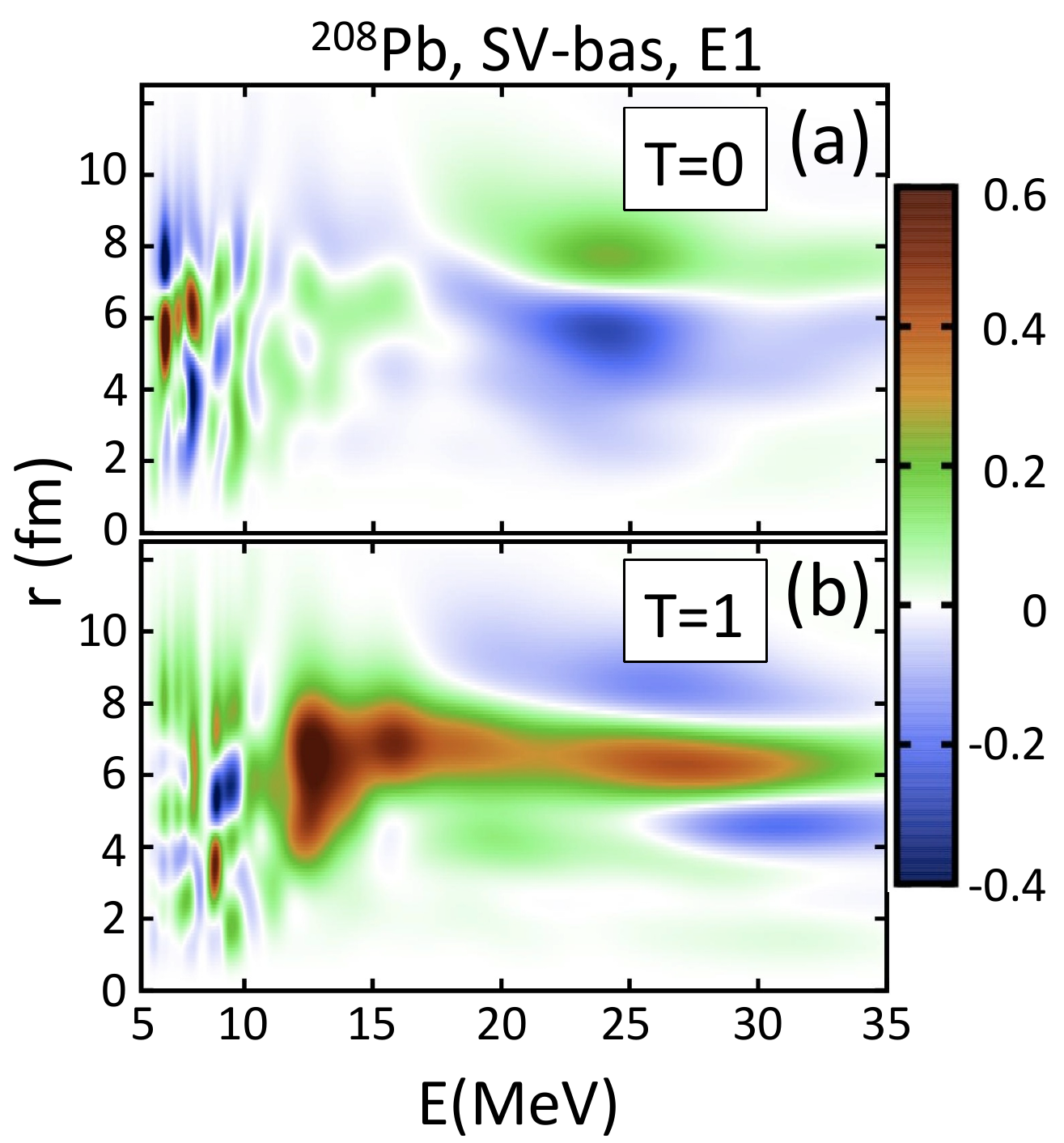}
  \caption[T]{\label{transdens}
  (Color online) Energy-averaged E1 transition density 
  in $^{208}$Pb  calculated with SV-bas as a function of $E$ and $r$ for $T = 0$ (top)
and $T = 1$ (bottom).
 }
\end{figure}

The interesting question is to what extent the lowest  E1 modes
can be
considered as collective, as  discussed (or assumed) in some studies.
To this end, it is instructive to study this point in the coordinate-space in terms of the 
transition density  $\rho_\nu^{(T)}$ of Eq.~(\ref{eq:rhotra}). In order to present results as a function of $E$, we performed a Gaussian folding as in Eq.~(\ref{eq:SFtra}). Figure~\ref{transdens} depicts the general behavior of E1 transition density  in $^{208}$Pb  as a function of $E$ and $r$.   At $E<12$\,MeV,
$\rho^{(T)}$   varies rapidly with $E$, exhibiting a complex multi-nodal behavior in both isospin channels (cf. discussion in Refs. \cite{(Pap11),*(Pap12),(Vre12)}). The strong state dependence, together with the  lack of a common pattern, are both  fingerprints of weak  collectivity. 
A different situation is seen in the GDR region. 
Although the energy band of  12-17 MeV covers a large
number of RPA  states, the $T=1$ transition density has the same pattern
throughout, with a pronounced surface bump. A similar pattern is seen for the 
$T=0$ compressional mode at 22-28 MeV.
Such coherent behavior is
characteristic of a
collective mode.

\section{Low-energy electric dipole strength}\label{LEE1strength}

\begin{figure}[htb]
  \includegraphics[width=0.8\columnwidth]{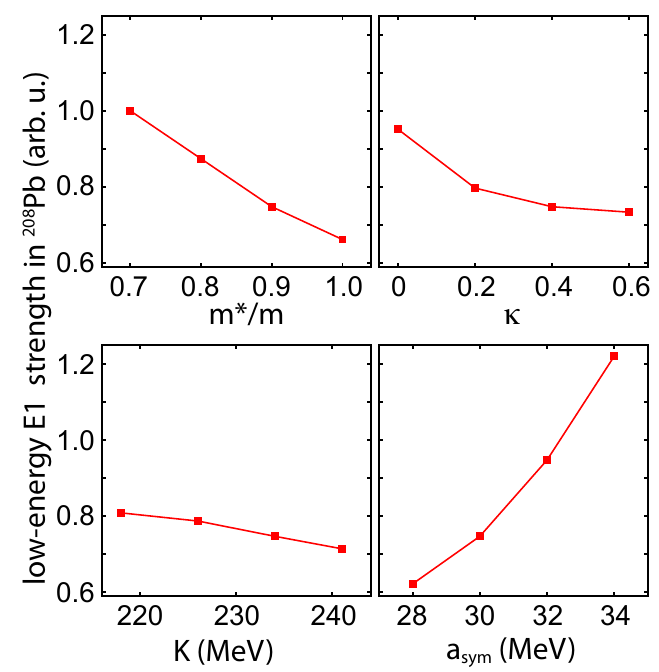}
  \caption[T]{\label{pygmycontrib}
  (Color online) Dependence of the low-energy E1 strength (\ref{PDR})
  on selected NMP ($K, m^*/m, a_\mathrm{sym}, \kappa$) in $^{208}$Pb predicted with SV-bas family of EDFs. The cutoff energy is $E_{\rm max}$=10\,MeV.
 }
\end{figure}

Although transition density and form factor reveal a world of
  local details, it is desirable to quantify their information content  in terms
  of a few integrated quantities. One such quantity  is the
accumulated
low-energy E1 strength, 
\begin{equation}\label{PDR}
B(E1; E_{\rm max})=\sum_{\nu\in\mathrm{RPA}, E_\nu<E_{\rm max}} B(E1,\nu), 
\end{equation}
which is proportional to the accumulated strength $A_{0}^D(E_{\rm max})$. 
This observable is somewhat ambiguous as it depends on  an assumed cutoff energy $E_{\rm max}$.
Figure~\ref{pygmycontrib} shows the predicted 
$B(E1; E_{\rm max})$ in $^{208}$Pb   as a function of two isoscalar 
(incompressibility $K$ and isoscalar effective mass $m^*/m$) and two isovector
(symmetry energy $a_\mathrm{sym}$ and TRK sum rule
enhancement $\kappa$) NMP indicators \cite{(Rei10)}.
The calculations were carried out with four sets of EDF parameterizations around SV-bas \cite{(Klu09)}  by systematically varying  the 
NMP of interest. We used the value 
 $E_{\rm max}$=10\,MeV that corresponds to the beginning of the GDR region in SV-bas, see Fig.~\ref{strengthforces}.

When considering each variation independently, one can see a monotonic trend in each case. But does it mean that a well-defined correlation exists between the low-energy E1 strength and the bulk properties considered? Clearly, the answer to this question cannot be given based on Fig.~\ref{pygmycontrib} alone, as $B(E1; E_{\rm max})$ seems to depend on several NMP, both isoscalar and isovector in character,  and the plot says nothing about the possible coupling between them as an underlying EDF parametrization is systematically varied. Second, if a parametrization
is too constrained in a given interaction  channel, {\it all}
variations probing this channel are correlated by construction. (An
example of such a situation was given in Ref.~\cite{(Rei10)} in the
context of a relativistic mean field model RMF-$\delta$-t, too
constrained  in the isovector channel to be used in a meaningful
correlation analysis.) 
Although the explicit tracking  with respect to
  single NMP, as done in Fig.~\ref{pygmycontrib}, can be very useful
  to uncover hidden dependencies between observables, it does not explore all
  conceivable variations and is thus insufficient to quantify
  true correlations. A more exhaustive method will
be discussed in  the next section.

\section{Correlations}\label{E1correlations}

Thus far we have analyzed the structure of dipole modes in terms of transition form factor and transition density. These structures do not show convincing signatures of collectivity in the low-energy region below GDR. As a more global measure, we inspected the integrated low-energy dipole strength, which indeed displays  some complex dependence on NMP. For a more exhaustive (and quantitative) analysis, we will now exploit the stringent method of covariance analysis and scrutinize the correlations between  integrated dipole observables and collective bulk properties (NMP, dipole polarizability, and neutron skin).
 
As already mentioned in section \ref{Themodel}, the empirical
adjustment of EDF parameters produces a nearly-parabolic $\chi^2$
landscape near the minimum, i.e.,
$\chi^2(\mathbf{p})\approx\chi^2_0+
\sum_{ij}{p}_i{p}_j\partial^2_{p_ip_j}\chi^2$.
The vicinity of the minimum in the $\mathbf{p}$-space, for which
$\chi^2(\mathbf{p})=\chi^2_0+1$, is considered as the space of
acceptable parameter variations ($1\sigma$ region). Observables are
also functions of the EDF parameters: $A=A(\mathbf{p})$.  A
correlation between observables $A$ and $B$ within a given model can
then be assessed by means of the correlation coefficient
\begin{equation}
{C}_{AB} =
\frac{|\overline{\Delta A\,\Delta B}|}
    {\sqrt{\overline{\Delta A^2}\;\overline{\Delta B^2}}},
\label{correlator}
\end{equation}
where the overline means an average over the space of acceptable
  parameters \cite{(Rei10),(Pie11),(Fat11)}.  Such correlation
  analysis had been proven useful in previous surveys.  For example,
it was concluded that the electric dipole polarizability is a good
isovector indicator that strongly correlates with the neutron radius
of $^{208}$Pb~\cite{(Rei10),(Pie11)}.

\begin{figure}[htb]
  \includegraphics[width=0.7\columnwidth]{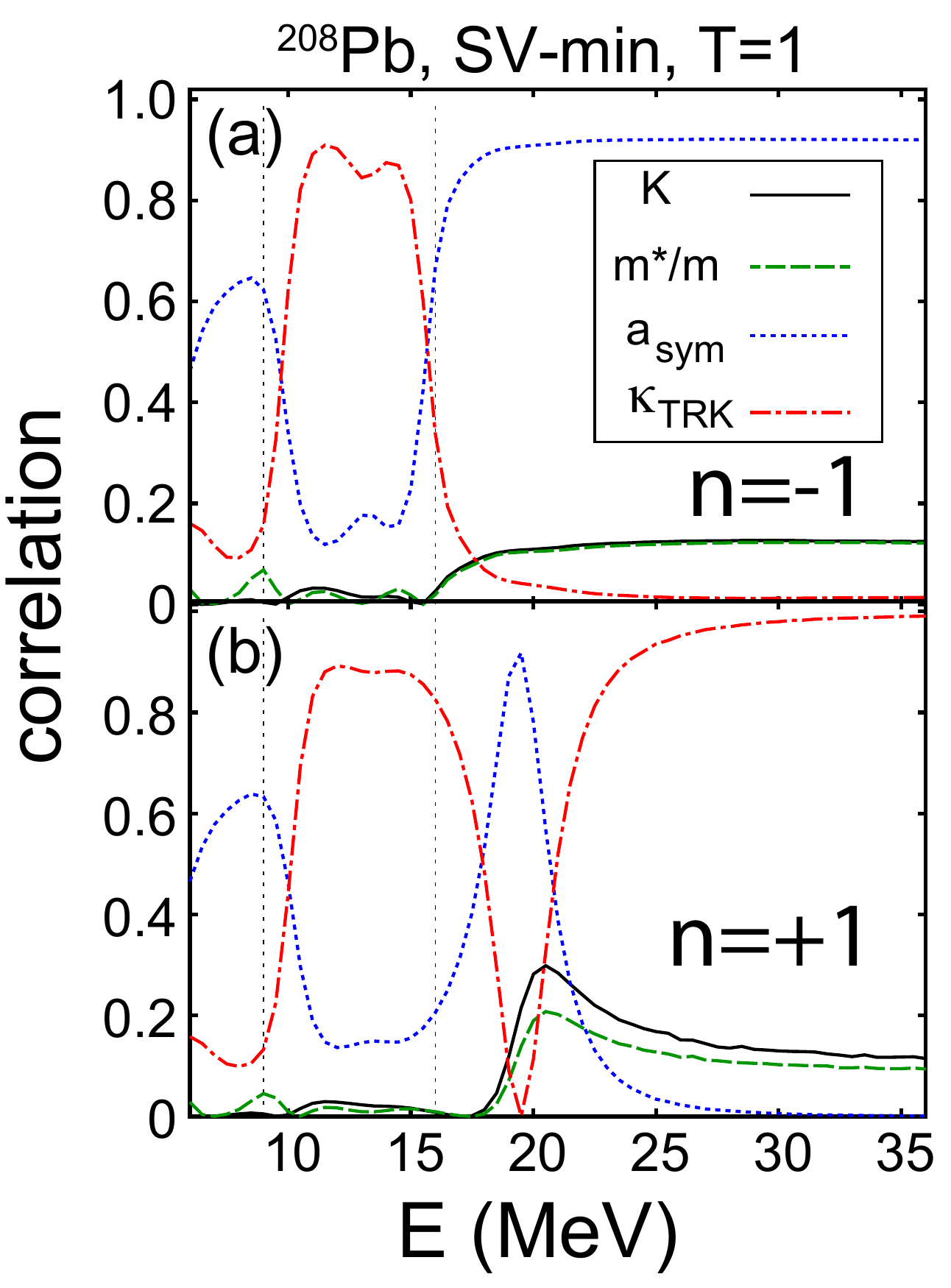}
  \caption[T]{\label{FEPM1-NM}
  (Color online) Correlation between the energy-weighted accumulated strength $A_n^D(E)$ (\ref{summedS})
  in $^{208}$Pb and selected NMP (incompressibility $K$, isoscalar effective mass $m^*/m$,
symmetry energy $a_\mathrm{sym}$, and TRK sum rule
enhancement $\kappa$) as a function of $E$ obtained with SV-min. Top: $n=-1$; bottom: $n=1$.
The GDR region is marked by vertical dotted lines.
 }
\end{figure}
\begin{figure}[htb]
  \includegraphics[width=0.7\columnwidth]{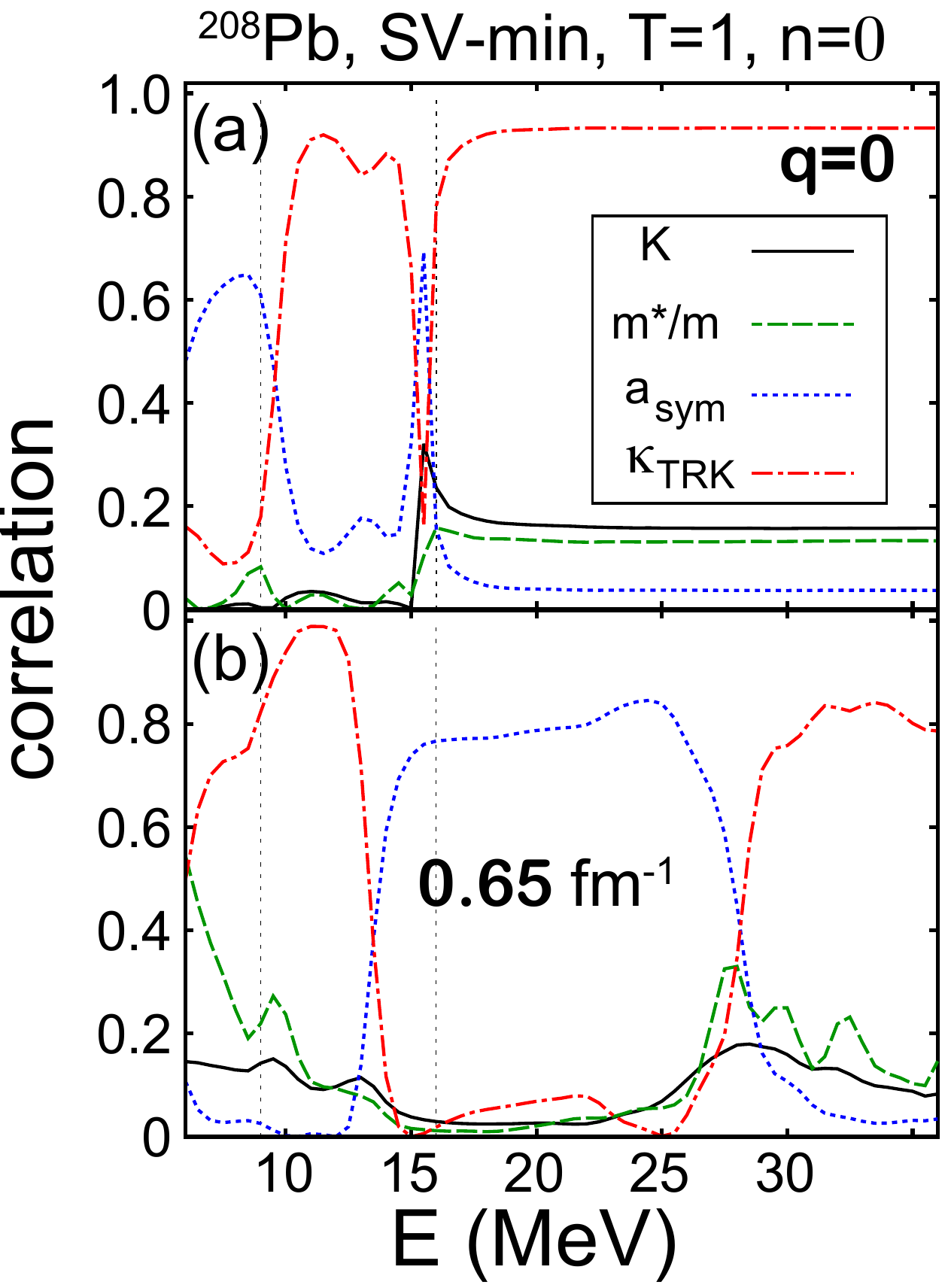}
  \caption[T]{\label{FE0T1-NM}
  (Color online) Similar to Fig.~\ref{FEPM1-NM} but for  the accumulated dipole  strength $A_0^D(E)$ 
  for two values of momentum transfer:  (a) $q=0$  and (b) $q=0.65$\,fm$^{-1}$.
 }
\end{figure}
\begin{figure}[htb]
  \includegraphics[width=0.7\columnwidth]{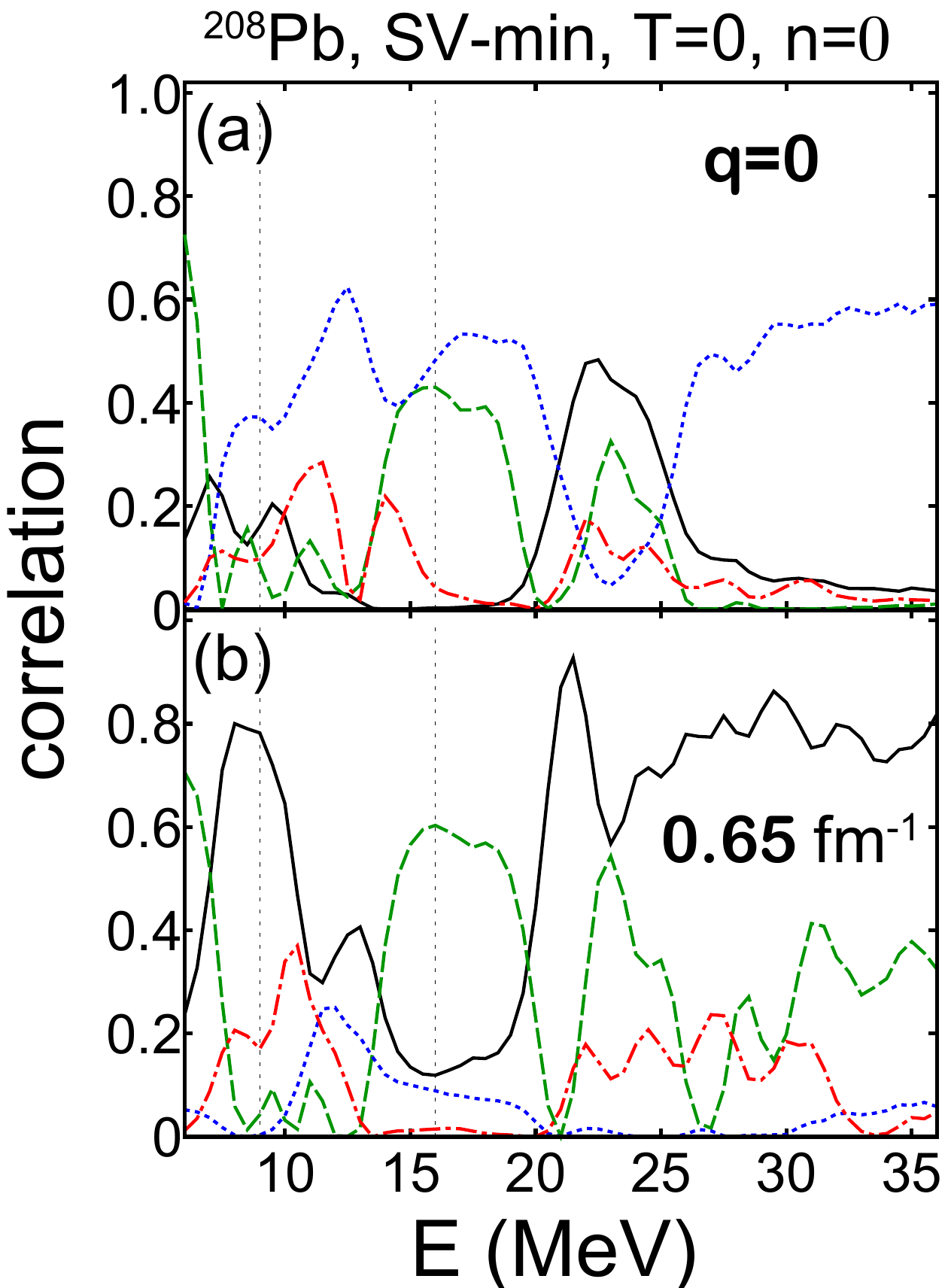}
  \caption[T]{\label{FE0T0-NM}
  (Color online) Similar as in Fig.~\ref{FE0T1-NM} but for
 the $T=0$  compressional dipole E1 strength.
 }
\end{figure}

The correlation between E1 strengths in $^{208}$Pb and
basic nuclear
matter properties ($K, m^*/m, a_\mathrm{sym}, \kappa$)   
is discussed in Figs.~\ref{FEPM1-NM}-\ref{FE0T0-NM} as a function of $E$.
Figure~\ref{FEPM1-NM} illustrates the correlation with 
the accumulated dipole strength $A_n^D(E)$ (\ref{summedS}).
The case of $n=-1$ is shown in Fig.~\ref{FEPM1-NM}(a).
As noted earlier, $A_{-1}^D(E)$ is the accumulator for the dipole polarizability $\alpha_{\rm D}$. It is interesting to see how the sensitivity to $a_\mathrm{sym}$ (and insensitivity to other NMP) develops at high energies. At GDR energies and below, however, the correlations vary dramatically. The changes are particularly 
rapid around the lower end of the GDR region (9-10\,MeV), which is often used as   a cutoff energy $E_{\rm max}$ to determine the low-energy E1 strength (\ref{PDR}).  It is apparent that even small changes in $E_{\rm max}$ may change correlations. This result casts serious doubts on standard (cutoff-dependent) definitions of  the pygmy strength.

The corollary of the emerging specific sensitivity is seen in
Fig.~\ref{FEPM1-NM}(b), which illustrates the case of $n=1$. At large energies, the accumulated
dipole strength $A_{1}^D(E)$ develops into an unambiguous measure of $\kappa$.
As expected, $K$  and $m^*/m$ never correlate with $A_{\pm 1}^D(E)$, regardless of the energy region. This is to be expected as these NMP are  isoscalar indicators \cite{(Rei10)}.  

The summed dipole  strength $A_0^D(E)$ is often used as a measure of the 
pygmy mode. We show  in
Fig.~\ref{FE0T1-NM} how it correlates with NMP. At zero momentum transfer, Fig.~\ref{FE0T1-NM}(a),
the correlation pattern strongly resembles that of Fig.~\ref{FEPM1-NM}(b), including strong correlation with $\kappa$ at high energies. The three cases displayed in Figs. \ref{FEPM1-NM} and \ref{FE0T1-NM}(a) have certain common features. Namely, they all show (i) rapid
changes near the lower and the upper end of the GDR region; (ii) strong correlation with $\kappa$ in the GDR region; and (iii) medium correlation with
$a_\mathrm{sym}$ around $E=7$\,MeV.

A stronger handle on $a_\mathrm{sym}$  is provided by the summed dipole strength
$A_0^D(E)$ at  nonzero momentum transfer. Figure \ref{FE0T1-NM}(b)
illustrates the case of integrated form factor strength at $q=0.65$\,fm$^{-1}$. A particularly large correlation with  $a_\mathrm{sym}$ is predicted in a wide energy range above the GDR region and below 25\,MeV. A strong correlation  with $\kappa$ in the GDR region still holds  at this value of $q$.

The complementing $n=0$ isoscalar strengths are shown in Fig.~\ref{FE0T0-NM}(a).
At low values of $q$, the  summed compressional dipole  strength $A_0^{D_c}(E)$ 
contains little information on NMP. The  sensitivity to $a_\mathrm{sym}$ at the low-$E$ region, suggested in \cite{(Vre12)}, is minor.
Figure~\ref{FE0T0-NM}(b) demonstrates that some sensitivity to   $K$    appears  at $q=0.65$\,fm$^{-1}$, especially at low energies below the GDR region and
in the high energy region above 20\,MeV.

\begin{figure}[htb]
  \includegraphics[width=0.7\columnwidth]{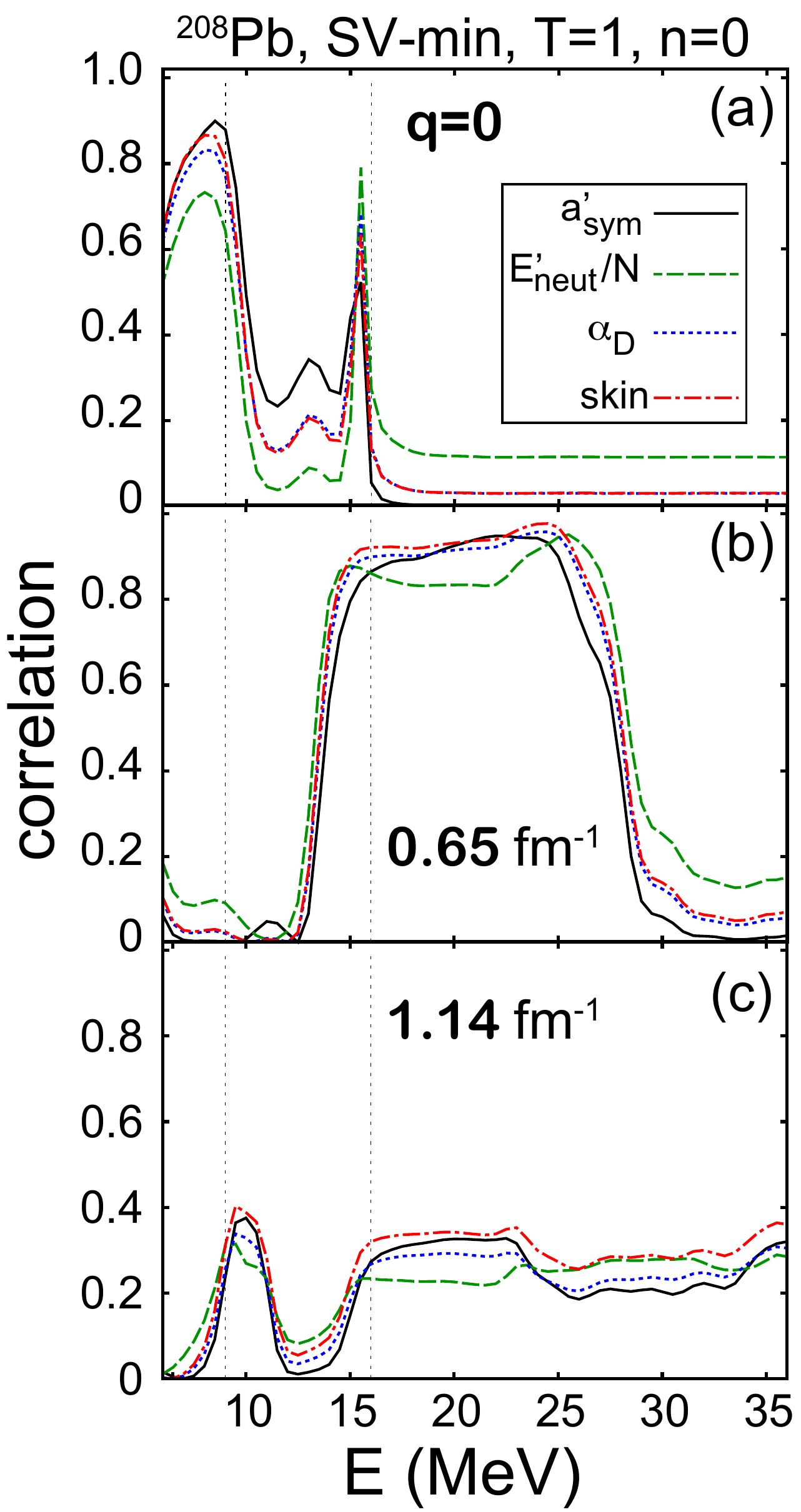}
  \caption[T]{\label{FE0corr}
  (Color online) Correlation between the $T=1$ accumulated E1 strength
  in $^{208}$Pb and selected isovector indicators ($a'_\mathrm{sym}, E′_\mathrm{neut}/N, \alpha_{\rm D}, r_{\rm skin}$)  in SV-min
  as a function of $E$ for $q=0$ (a);  $q=0.65$\,fm$^{-1}$ (b); and $q=1.14$\,fm$^{-1}$ (c).
 }
\end{figure}

We now turn to a correlation between isovector form factor strength $A_0^F(E)$ and the four strong isovector indicators: 
the slope of symmetry energy $a'_\mathrm{sym}$ at the saturation density,
slope of the neutron EoS $E′_\mathrm{neut}/N$
at neutron density 0.08\,fm$^{−3}$, E1 polarizability $\alpha_{\rm D}$, and neutron-skin thickness $r_{\rm skin}$ in $^{208}$Pb.
Recall that these four observables are highly
correlated with each other and with $a_\mathrm{sym}$ \cite{(Rei10)}.
Consequently, all four correlations displayed in Fig.~\ref{FE0corr} are practically identical,
and also agree  with the results for $a_\mathrm{sym}$ shown in Fig.~\ref{FE0T1-NM}: the summed isovector E1 strength above the GDR at  the intermediate values of momentum transfer around  $q=0.65$\,fm$^{-1}$ is an excellent isovector indicator. As illustrated  in Fig.~\ref{FE0corr}(c), however,  this correlation deteriorates at higher values of $q$.

\begin{figure}[htb]
  \includegraphics[width=0.7\columnwidth]{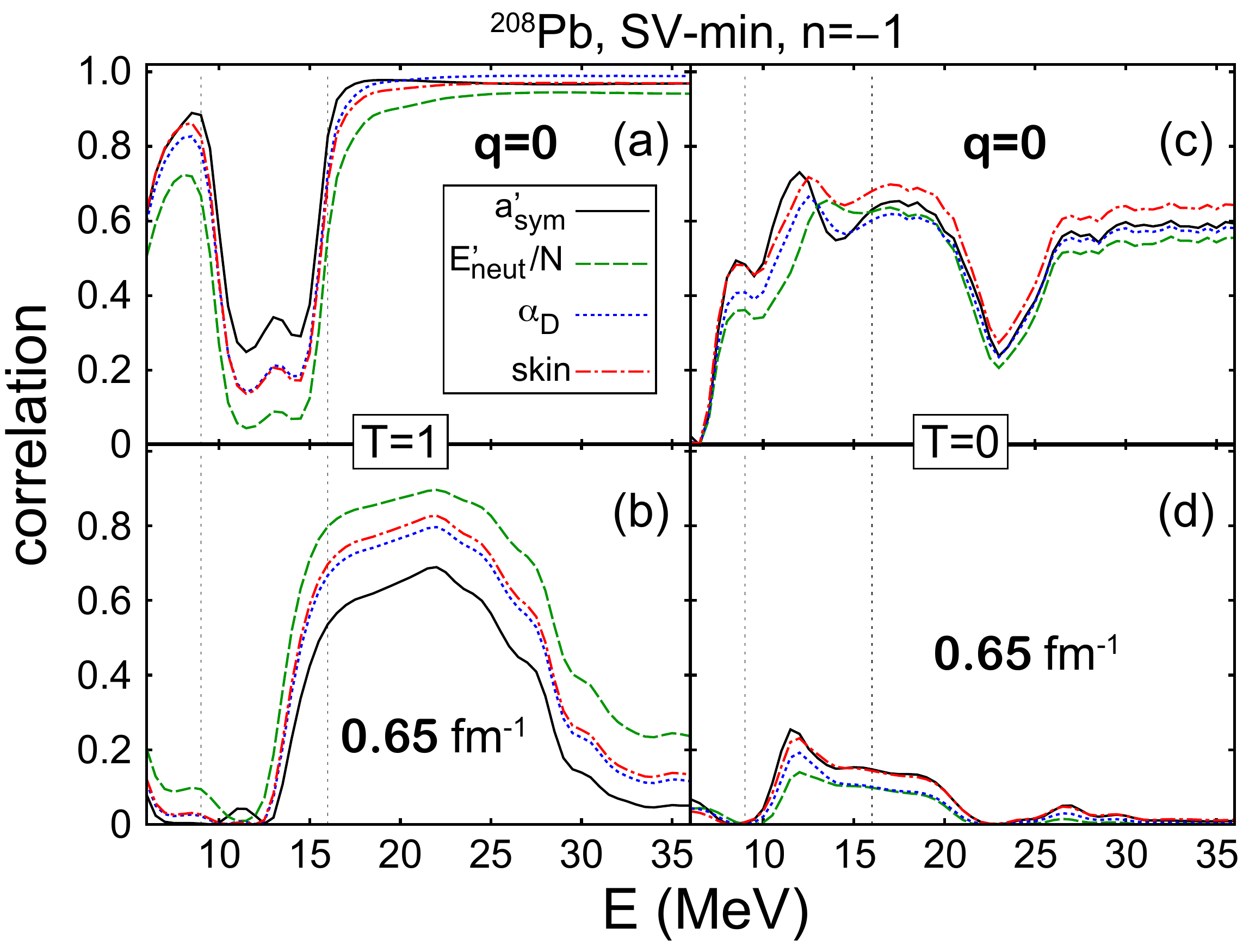}
  \caption[T]{\label{FEM1corr}
  (Color online) Correlation between the $T=1$ (left) and $T=0$ (right) accumulated inverse-energy-weighted  E1 strength $A_{-1}^F$ 
  in $^{208}$Pb and selected isovector indicators ($a'_\mathrm{sym}, E′_\mathrm{neut}/N, \alpha_{\rm D}, r_{\rm skin}$)  in SV-min as a function of $E$ for $q=0$ (top) and   $q=0.65$\,fm$^{-1}$ (bottom).
 }
\end{figure}

Fig.~\ref{FEM1corr} illustrates the  information content
of the  accumulated inverse-energy-weighted E1  strength $A_{-1}^F(E)$ with respect to the four isovector observables discussed in the context of Fig.~\ref{FE0corr}.
Again, we see that the correlations are very similar for all four observables and
the $T=1$ accumulated  strength at $q=0.65$\,fm$^{-1}$ nicely correlates with isovector indicators around $E=20$\,MeV. The accumulated inverse-energy-weighted isoscalar strength does not seem to correlate well with isovector observables.

\section{A quick glance at $^{132}$S\lowercase{n}}\label{Sncase}

\begin{figure}[htb]
  \includegraphics[width=0.7\columnwidth]{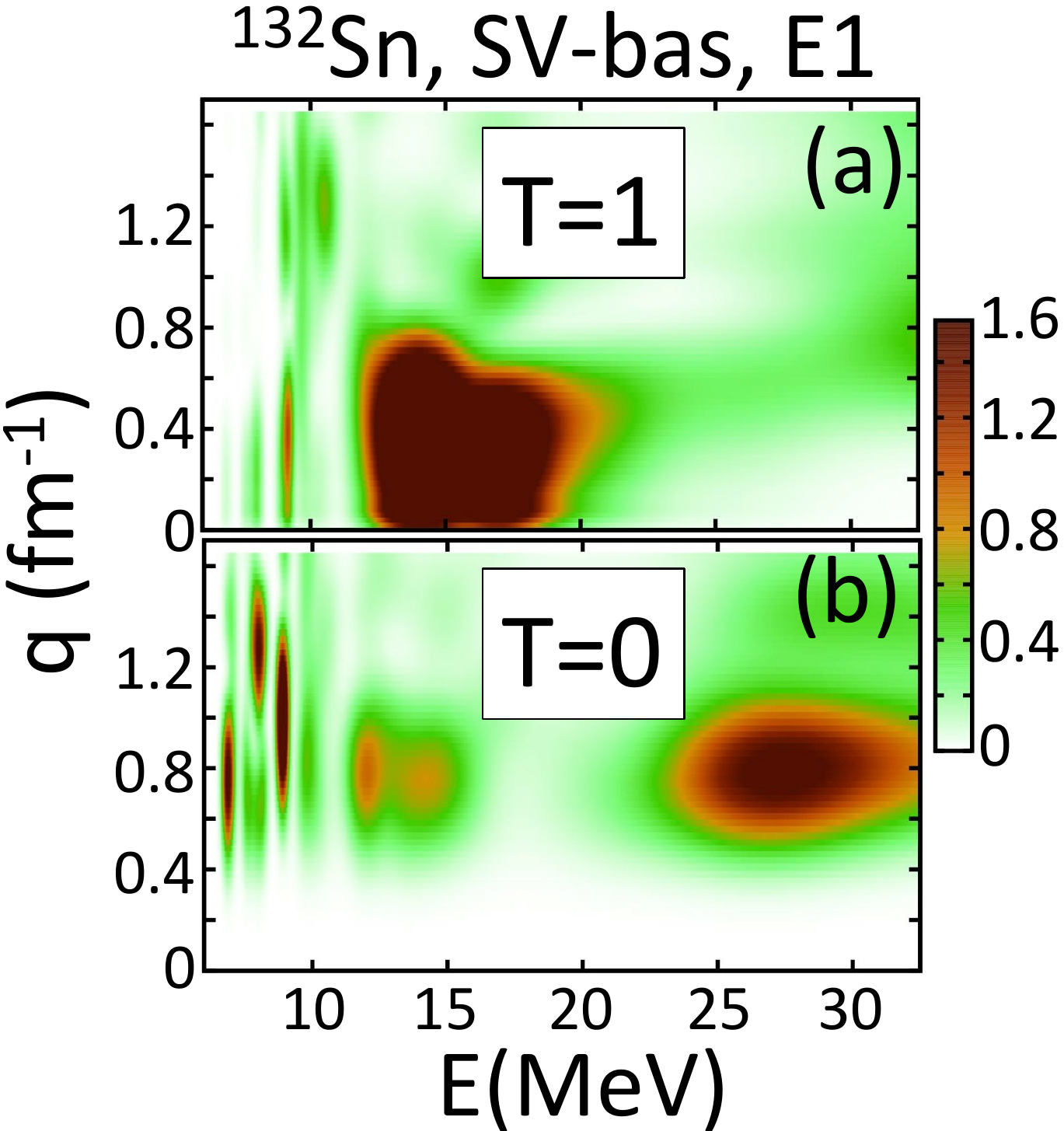}
  \caption[T]{\label{formfactors-132Sn}
  (Color online) Similar as in Fig.~\ref{formfactors}, panels  (c) and (d),  except  for $^{132}$Sn.
 }
\end{figure}
It is interesting to check whether the  findings obtained for $^{208}$Pb
are specific to this nucleus or whether they are of a more general
nature. To this end, we inspect the case of a neutron-rich,
doubly-magic nucleus $^{132}$Sn. Figure~\ref{formfactors-132Sn} shows
its  $T=1$ and $T=0$ E1 strength  predicted with SV-bas. The overall pattern,
with GDR transitions in the isovector channel
and the dominant dipole compressional mode in the isoscalar channel,
closely resembles that shown in  Fig~\ref{formfactors}
for $^{208}$Pb. In particular,
the low-energy modes are isoscalar  at low-$q$ values and isoscalar at higher momentum transfer. 

\begin{figure}[htb]
  \includegraphics[width=0.7\columnwidth]{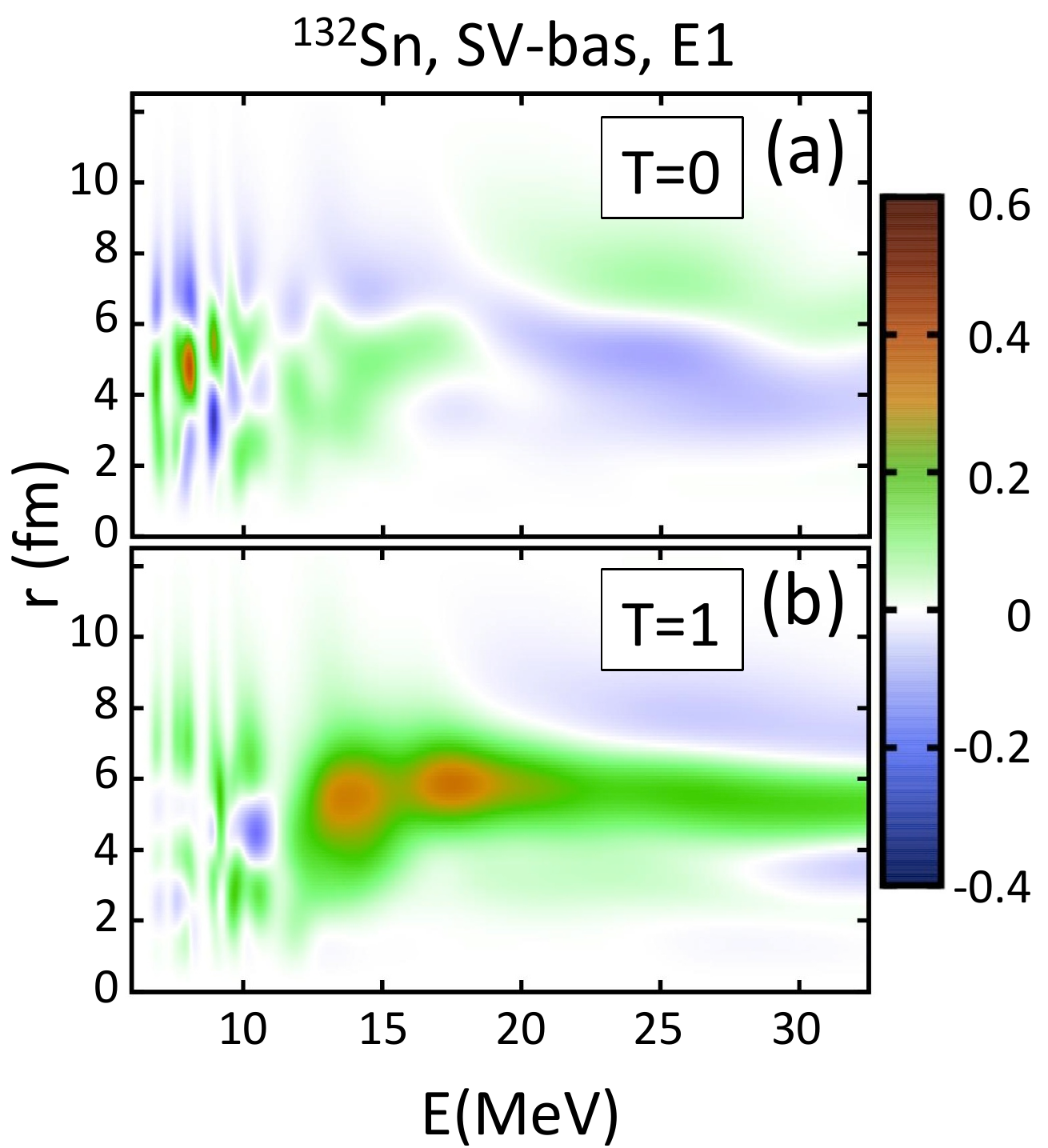}
  \caption[T]{\label{transdens-132Sn}
  (Color online) Similar as in Fig.~\ref{transdens}, except for $^{132}$Sn.
 }
\end{figure}
\begin{figure}[htb]
  \includegraphics[width=0.7\columnwidth]{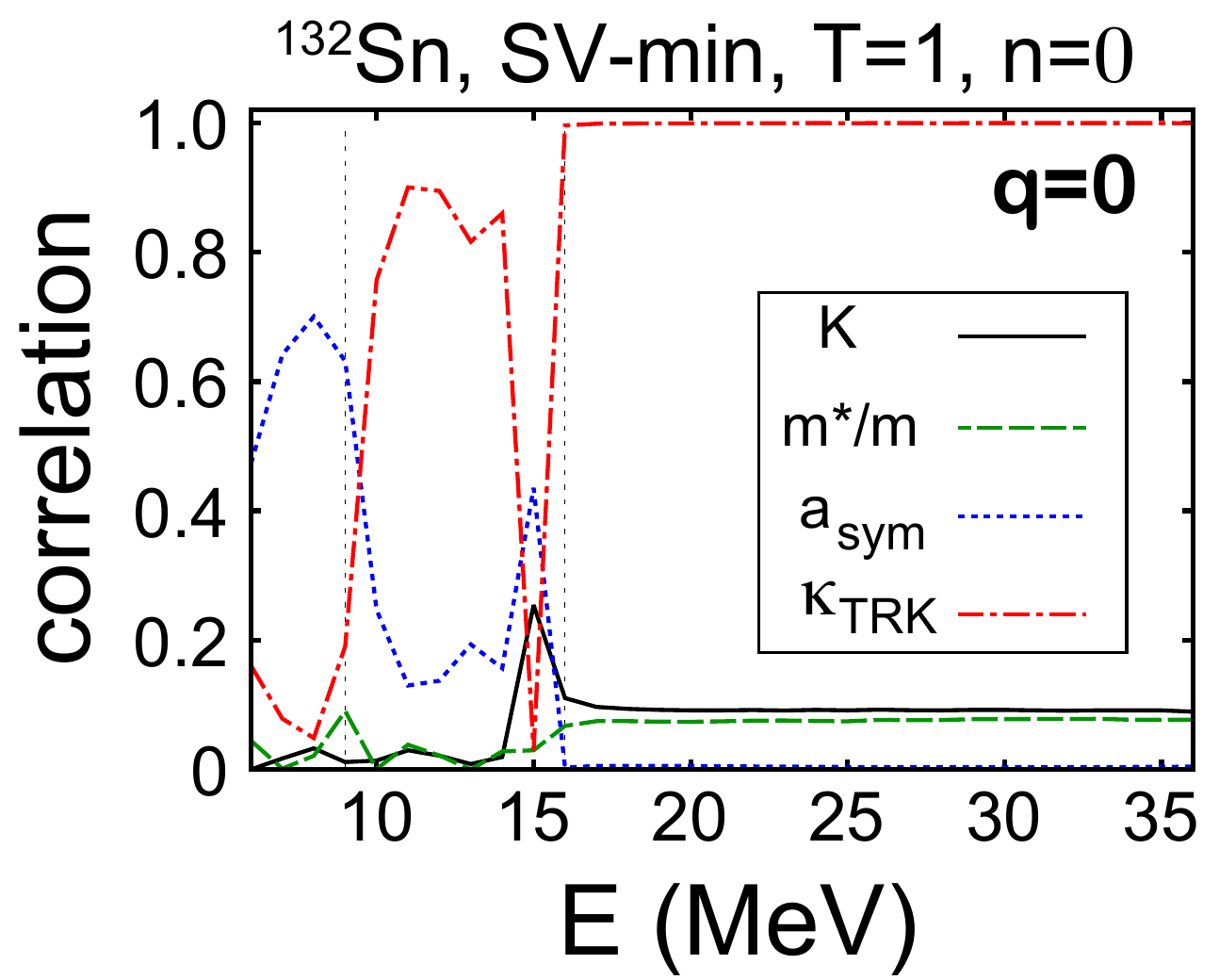}
  \caption[T]{\label{FE0T1-NM-132Sn}
  (Color online)  Similar as in Fig.~\ref{FE0T1-NM}(a), except for $^{132}$Sn.
 }
\end{figure}

Figure~\ref{transdens-132Sn} shows the transition densities for
$^{132}$Sn. Again, the pattern is extremely similar to the case of
$^{208}$Pb. Finally Fig.~\ref{FE0T1-NM-132Sn} shows the correlations of the four basic NMP
with accumulated isovector dipole strength. It is, again, similar to the
analogous Fig.~\ref{FE0T1-NM}(a) for $^{208}$Pb.
We also carried out calculations for other neutron-rich nuclei, such as $^{140}$Sn, and  our conclusion is that the case of $^{208}$Pb is representative to other spherical closed-shell  heavy
systems.

\section{Conclusions}\label{Conclusions}

To clarify the ambiguous theoretical situation with regard to the  interpretation of the low-energy electric dipole strength, we study its
 information content with respect to isovector
and isoscalar indicators characterizing bulk nuclear matter and finite nuclei. 
As a theoretical tool, we use self-consistent nuclear density functional theory 
augmented by the random phase approximation. The inter-observable correlations are computed by means of  covariance analysis.
We use  well-calibrated Skyrme energy density functionals that allow
for systematic variations of isoscalar and isovector parameters
characterizing nuclear matter properties. 
(The extension of the correlation analysis 
to other models, including relativistic EDF,  will be carried out in  a forthcoming study \cite{(Paa12)}.)
To avoid ambiguities due to
strong shell effects and pairing, our investigations have been limited 
to the doubly-magic heavy nuclei $^{208}$Pb and
on $^{132}$Sn, and the results were very close in both cases.
To separate  various electric dipole
modes, we study the E1 strength as a function of excitation
energy $E$ and  momentum transfer $q$. By going to the
$q$-dimension, we are able to resolve excitation modes that are close
in energy, and can nicely separate low-energy excitations from the $T = 1$ and $T = 0$ giant  resonances.

Our study fully confirms the main conclusion of our previous work
\cite{(Rei10)}. Namely, the low-energy dipole excitations cannot be
interpreted in terms of a collective pygmy dipole resonance  mode
associated with the motion of skin neutrons.  The detailed inspection of transition form factor and density shows a pattern of rapidly varying  particle-hole excitations. Moreover, it seems that a large fraction of the low lying states are isovector ``shadows'' of the underlying low-energy part of the isoscalar compressional mode.

At the level of more global observables, we have studied the E1 strength integrated up to a given cutoff energy. As a first indicator of correlations, we have plotted in Fig.~\ref{pygmycontrib} the changes of the low-energy  strength   with systematically varied nuclear matter properties. This has led to a complex picture where every NMP has some influence on the low energy strength. 
To assess  correlations in a meaningful way, we went beyond the simple inspection of trends and  employed the standard tools of statistical analysis. While the local E1 strength at low energies weakly correlates with isovector indicators such as symmetry energy, the magnitude of those correlations varies rapidly  around the lower end of the GDR region. This strong dependence on the cutoff energy indicates that a pygmy strength  cannot be unambiguously defined; hence, it is not a particularly useful quantity.

Of particular importance are the fully integrated strength; the
  much celebrated sum rules. The $T=1$ sum rules, such as the dipole
polarizability $\alpha_\mathrm{D}$ (the inverse energy weighted
  sum rule), the sum rule enhancement factor $\kappa$ (strongly
  correlated with the energy weighted sum rule), and the total
accumulated E1 strength are all excellent probes of nuclear
isovector properties.  At the isoscalar side, we found a
relatively strong correlation between nuclear incompressibility and
the accumulated isoscalar dipole strength for $q=0.65$\,fm$^{-1}$ (at
$E=9$\,MeV and above 20\,MeV). In general, experimental studies at $q
>0$ in the isovector and isoscalar channel could provide an excellent window for a better characterization of
E1 strength and relating it to fundamental nuclear properties.

\begin{acknowledgements}
This work was initiated during the Workshop on {\it The Nuclear Dipole Polarizability and its Impact on Nuclear Structure and Astrophysics},
held at ECT* in Trento, June 18-22, 2012. 
This work was supported by BMBF under contract no. 06~ER~142D
and by the Office of Nuclear Physics,  U.S. Department of Energy under Contract
No. DE-FG02-96ER40963.
\end{acknowledgements}


%

\end{document}